\documentclass[iop,apj,tighten,a4paper]{emulateapj}
\usepackage{amsmath,amstext,amssymb}
\usepackage[breaklinks,colorlinks,citecolor=blue,linkcolor=magenta]{hyperref}
\usepackage{epsfig}
\usepackage{appendix}
\usepackage{wrapfig}
\usepackage{times}
\usepackage{BibDef}
\usepackage{hyperref}
\usepackage{amssymb}
\usepackage{float}
\usepackage{breakurl}
\usepackage{color}

\def\spose#1{\hbox to 0pt{#1\hss}}
\def\lta{\mathrel{\spose{\lower 3pt\hbox{$\mathchar"218$}}
     \raise 2.0pt\hbox{$\mathchar"13C$}}}
\def\gta{\mathrel{\spose{\lower 3pt\hbox{$\mathchar"218$}}
     \raise 2.0pt\hbox{$\mathchar"13E$}}}
\def\HI{\hbox{\rm H~$\scriptstyle\rm I$}}

\def\HII{\hbox{\rm H~$\scriptstyle\rm II$}}
\def\nHI{{\rm HI}}
\def\nH{{\rm H}}
\def\nHII{{\rm HII}}

\def\nHeI{{\rm HeI}}
\def\nHeII{{\rm HeII}}
\def\nHeIII{{\rm HeIII}}
\def\HeI{\hbox{He~$\scriptstyle\rm I$}}
\def\HeII{\hbox{He~$\scriptstyle\rm II$}}
\def\HeIII{\hbox{He~$\scriptstyle\rm III$}}
\def\HeIII{\hbox{He~$\scriptstyle\rm III$}}

\def\kmsmpc{\,{\rm km\,s^{-1}\,Mpc^{-1}}}
\def\cmm{\,{\rm cm^{-2}}}
\def\cmc{\,{\rm cm^{-3}}}

\def\lumdens{\,{\rm erg\,s^{-1}\,Mpc^{-3}\,Hz^{-1}}}

\def\xiunits{\,{\rm erg^{-1}\,Hz}}

\def\Lya{Ly$\alpha$}

\lefthead{Madau}
\righthead{Cosmic Reionization after Planck and before JWST}
\submitted{}


\begin{document}

\title[Cosmic Reionization After Planck and Before JWST]{Cosmic Reionization After Planck and Before JWST: An Analytic Approach}

\author{Piero Madau\altaffilmark{1,2}}
\affil{$^1\,$Department of Astronomy \& Astrophysics, University of California, 1156 High Street, Santa Cruz, CA 95064, USA\\
$^2\,$Institut d'Astrophysique de Paris, Sorbonne Universit{\'e}s, UPMC Univ Paris 6 et CNRS, UMR 7095, 98 bis bd Arago, 75014 Paris, France\\
}

\begin{abstract}
The reionization of cosmic hydrogen marks a critical juncture in the history of structure formation in the universe. Here we present a new 
formulation of the standard reionization equation for the time evolution of the volume-averaged \HII\ fraction that is more consistent 
with the accepted conceptual model of inhomogeneous intergalactic absorption. The revised equation retains the basic terminology 
and simplicity of the classic calculation but explicitly accounts for the presence of the optically thick ``Lyman-limit systems" that are known
to determine the mean free path of ionizing radiation after overlap. Integration of the modified equation provides a better
characterization of the timing of reionization by smoothly linking the pre-overlap with the post-overlap phases of such process. We confirm 
the validity of the quasi-instantaneous approximation as predictor of reionization completion/maintenance, and discuss new insights on the 
sources of cosmic reionization using the improved formalism. A constant emission rate into the intergalactic medium (IGM) of about 
3 Lyman continuum (LyC) photons per atom per Gyr leads to a reionization history that is consistent with a number of observational constraints on the 
ionization state of the $z=$ 5--9 universe and with the reduced Thomson scattering optical depth recently reported by the Planck Collaboration.  
While star-forming galaxies can dominate the reionization process if the luminosity-weighted fraction of LyC photons that escape into the IGM, 
$f_{\rm esc}$, exceeds 15\% (for a faint magnitude cut-off of the galaxy UV luminosity function of $M_{\rm lim}=-13$ and a LyC photon yield per unit 
1500\,\AA\ luminosity of $\xi_{\rm ion}=10^{25.3}\,\xiunits$), simple models where the product of the two unknowns $f_{\rm esc}\xi_{\rm ion}$ is 
not evolving with redshift fail to reproduce the changing neutrality of the IGM observed at these epochs.
\end{abstract}
\keywords{cosmology: theory --- dark ages, reionization, first stars --- diffuse radiation --- intergalactic medium --- galaxies}

\section{Introduction}

The transformation of cold neutral intergalactic hydrogen into a highly ionized warm plasma marks the end of the cosmic dark ages and  
the beginning of the age of galaxies. Studies of resonant absorption in the spectra of distant quasars have shown that hydrogen reionization was 
still ongoing at $z\sim 6$ and fully completed by redshift 5.5 \citep[e.g.,][]{becker15,mcgreer15,fan06}. 
An early onset of reionization appears to be strongly disfavoured by the latest {\it Planck} analysis: the combination of CMB temperature anisotropies
with low-multipole polarization data yields an integrated Thomson scattering optical depth, $\tau_{\rm es}=0.058\pm 0.012$ \citep{planck16}, 
which is lower and more precise than previous measurements. The ionization state of the $z=$ 6--8 universe
is being constrained by other tracers of reionization history, from the damping wing absorption profiles in the spectra of quasars \citep[e.g.,][]{schroeder13}
to the luminosity function (LF) and clustering properties of \Lya\ emitting galaxies (LAEs) \citep[e.g.,][]{schenker14,ouchi10}. Such studies indicate that the 
intergalactic medium (IGM) was significantly neutral at redshifts between 6 and 7, in agreement with the {\it Planck} results.

While a broad consensus on the epoch and duration of the reionization process may be emerging from a variety of astrophysical probes, many key aspects, 
such as the very nature of the first sources of UV radiation, how they interacted with their environment, and the early thermodynamics of primordial baryonic gas, 
remain uncertain despite a considerable community effort over the last two decades \citep[for recent reviews on this topic see][]{haiman16,lidz16}.
A detailed modeling of reionization is very challenging, as it requires cosmological numerical simulations that self-consistently couple all the relevant 
physical processes -- dark matter dynamics, gas dynamics, self-gravity, star formation/feedback, radiative transfer, nonequilibrium ionization/recombination, 
chemical enrichment, heating and cooling -- over a large range of scales \citep[e.g.,][]{gnedin14,so14}.  
To zeroth order, however, understanding reionization is mainly about a proper accounting of the production and absorption of ionizing LyC
radiation as a function of epoch in a clumpy, expanding medium, and this is commonly achieved by solving a ``reionization equation" of the form \citep{madau99} 
\begin{equation}
{dQ\over dt} = {\langle \dot n_{\rm ion}\rangle \over \langle n_\nH \rangle}-{Q\over \bar t_{\rm rec}}
\\[8pt]
\label{eq:intro}
\end{equation}
for the time evolution of the volume-averaged hydrogen ionized fraction $Q$. Here, $\langle {\dot n}_{\rm ion} \rangle$ is the emission rate into the 
IGM of ionizing photons per unit proper volume, $\langle n_\nH\rangle=1.89\times 10^{-7}\,(1+z)^3\,\cmc$ is the cosmological mean proper density of hydrogen, and 
$\bar t_{\rm rec}$ is an effective recombination timescale. It is this simple ODE that statistically describes the transition from a neutral universe to a fully 
ionized one, turning reionization into a photon-counting exercise in which the growth rate of \HII\ regions is equal to the rate at which ionizing photons are
produced minus the rate at which they are consumed by radiative recombinations in the IGM \citep{madau99}. Extensively used in the literature 
\citep[e.g.,][]{haardt12,kuhlen12,bouwens15planck,robertson15,madau15,khaire16,ishigaki17,sharma17} as it allows an estimate of the 
photon budget required to achieve reionization with a fast exploration of parameter space, Equation (\ref{eq:intro}) has been shown to provide a good 
description of the results of radiative transfer simulations \citep{gnedin00,gnedin16}. With one major limitation: it assumes that all 
LyC photons escaping from individual galaxies are absorbed by the diffuse IGM, mathematically permitting values of $Q$ that are above unity when 
reionization is completed and \HII\ regions overlap, which is physically impossible.

At $z<6$, however, the diffuse IGM is observed to be highly ionized, with only a small residual amount of neutral hydrogen set by the balance between radiative 
recombinations and photoionizations from a nearly uniform UV radiation background, and provides negligibly small \HI\ photoelectric absorption. The continuum opacity 
is instead dominated by the optically thick (to ionizing radiation) ``Lyman-limit systems" (LLSs), high density regions that trace non-linear and collapsed structures, 
occupy a small portion of the volume, and are able to keep a significant fraction of their hydrogen in neutral form \citep[e.g.][]{miralda00,gnedin06,furlanetto09,haardt12}.
In this paper we present a new formulation of the standard reionization equation that explicitly accounts for the presence of these LLSs. 
We shall see how the LLS opacity causes the mean free path of LyC radiation to remain relatively small even after overlap, and $Q$ never exceeds unity. 
The revised ODE retains the basic terminology and simplicity of the classic calculation but is more consistent with the accepted conceptual model of inhomogeneous 
intergalactic absorption, and provides a better characterization of the timing of reionization by connecting the 
pre-overlap with the post-overlap epochs. We use the improved formalism for a fresh reassessment of a scenario
in which star-forming galaxies dominate the production of LyC radiation in the pre-overlap, overlap, and immediate post-overlap stages of cosmic hydrogen reionization.

Below, we shall adopt a $(\Omega_M,\Omega_\Lambda,\Omega_b)=(0.3,0.7,0.045)$ flat cosmology with $H_0=70\,\kmsmpc$. The hydrogen and helium mass fractions are $X=0.75$ and 
$Y=0.25$, respectively.

\section{Basic Formalism}
We start by summarizing the basic theory describing the impact of ionizing radiation on an inhomogeneous, primordial IGM.
The general idea of a gradual reionization process driven by a steadily increasing UV photon production rate can be cast into a quantitative framework 
by integrating the rate equations for the fractional abundances of the three species \HII, \HeI, and \HeIII, 
\begin{align}
\label{eq:xHII}
{d\langle x_\nHII\rangle \over dt} & = \langle x_\nHI \Gamma_\nHI\rangle - \langle n_e x_\nHII \alpha_\nHII\rangle, \\ 
\label{eq:xHeI}
{d\langle x_\nHeI\rangle \over dt} & = -\langle x_\nHeI \Gamma_\nHeI\rangle + \langle n_e x_\nHeII \alpha_\nHeII\rangle,\\
\label{eq:xHeIII}
{d\langle x_\nHeIII\rangle \over dt} & = \langle x_\nHeII \Gamma_\nHeII\rangle - \langle n_e x_\nHeIII \alpha_\nHeIII\rangle,
\end{align}
supplemented with three closure conditions for the conservation of charge and of the total abundances of hydrogen and helium.
Here, $\Gamma_i$ is the photoionization rate of ion $i$, $\alpha_j(T)$ is the radiative recombination coefficient (in units of volume per unit time) of species $j$ into $i$,
$n_e=n_\nHII+n_\nHeII+2n_\nHeIII$ is the proper electron density, and the angle brackets denote an average over all space.
The evolution of, e.g., the quantity $\langle x_\nHII\rangle$ with redshift fully specifies the global reionization history of cosmic hydrogen.
The volume-averaged hydrogen photoionization rate can be written as 
\begin{equation}
\langle \Gamma_\nHI\rangle = \int d\nu\, \frac{4\pi J_\nu}{h\nu}\,\sigma_\nHI(\nu), 
\\[8pt]
\end{equation}
where $\sigma_\nHI$ is the photoionization cross-section and $J_\nu(t)$ is the mean monochromatic UV radiation intensity, averaged over all space and directions,
 \begin{equation}
J_\nu(t)\equiv \lim_{V\to\infty} ~ {1\over 4\pi V}\int_V d^3x \int d\Omega\, I_\nu(t,\vec{x},\hat{n}).
\\[8pt]
 \end{equation}
Here, $I_\nu(t,\vec{x},\hat{n})$ is the specific intensity of the radiation field at time $t$ and position $\vec{x}$, measured along the direction $\hat{n}$. 
The integral solution of the equation of radiative transfer is given by
\begin{equation}
J_\nu(t)={c\over 4\pi}\int_0^{t}\, {dt'} h\nu' \langle {\dot n}_{\nu'}(t')\rangle \left[{a(t')\over a(t)}\right]^3e^{- {\bar \tau}(\nu,t,t')}, 
\\[8pt]
\label{Jnu}
\end{equation}
where $\nu'=\nu a(t)/a(t')$, $\nu''=\nu a(t)/a(t'')$, $\langle {\dot n}_{\nu}\rangle$ is the specific photon emission rate into the IGM per unit proper volume, $a(t)$ is the 
cosmic scale factor, and
\begin{equation}
\bar \tau(\nu,t,t')=c\int_{t'}^{t} dt'' \bar \kappa_{\nu''}(t'')
\\[8pt]
\label{taueff}
\end{equation}
is an ``effective optical depth" for photons traveling from $t'$ to $t>t'$.  A packet of photons will travel a proper mean free 
path $1/\bar \kappa_\nu$ before suffering an $1/e$ attenuation. 

The mean opacity $\bar \kappa_\nu$ in Equation (\ref{taueff}) is defined as $\bar \kappa_\nu\equiv \langle \kappa_\nu I_\nu\rangle/J_\nu$, where the average is taken 
over all space and all directions \citep[e.g.,][]{gnedin97}. This is not the {\it space average} of the absorption coefficient, since it is weighted in the radiative 
transfer equation by the local value of the specific intensity $I_\nu$. The relation $\bar \kappa_\nu=\langle \kappa_\nu\rangle$ holds only in the limit of a 
uniform and isotropic ionizing background, or when $\kappa_\nu$ and $I_\nu$ are independent random variables.
This is a reasonable approximation after overlap (the point at which \HII\ regions merge and the ionizing background rises by a large factor)
or during the late stages of reionization, when ionized bubbles expand into low-density regions under the
collective influence of many UV sources \citep[e.g.,][]{iliev06}, the photon mean free path is insensitive to the strength of the local radiation field,
and variations in the LyC opacity are relatively modest. The assumption of a nearly uniform UV background, however, breaks down during the early stages 
of reionization, when rare peaks in the density field that contain more absorbers as well as more sources ionize first, and the photon mean free path is 
comparable to or smaller than the average source separation.
A detailed modeling of the spatial correlations between sources and sinks of ionizing radiation 
can only be achieved using radiative transfer simulations \citep[e.g.,][]{gnedin14,so14} or semi-numerical schemes based on the excursion set 
formalism \citep[e.g.][]{furlanetto04,mesinger11,zahn11,kaurov13}.
It is possible to capture some of these complexities into a single parameter, the ``photoionization clumping factor" $C_I$, that accounts for the 
cross-correlation between the fractional abundance of \HI\ and the radiation field, and write 
\begin{equation}
\bar \kappa_\nu=C_I \langle \kappa_\nu\rangle.
\\[8pt]
\end{equation}
In reionization studies it is conventional to include in the ionization balance equation only a fraction of the hydrogen photoionizations and radiative 
recombinations, i.e. those that take place in the low-density IGM. The rest are absorbed in-situ, in high-density regions within virialized halos, close 
to the ionizing sources. As shown by \citet{kohler07}, when all local ionizations and absorptions are removed from cosmological simulations, the photoionization
clumping factor $C_I$ is of order unity close to overlap, and decreases at earlier times as the neutral fraction is anticorrelated with the radiation field.

When the photon mean free path is much smaller than the horizon size, $1/\bar \kappa_\nu\ll c/H$, cosmological effects such as source evolution 
and frequency shifts can ne neglected. In this local, $\bar\tau\rightarrow \infty$  approximation, the mean specific intensity relaxes to the 
{source function} $\langle{\dot n}_\nu\rangle/\bar \kappa_\nu$, 
\begin{equation}
4\pi J_\nu\simeq h\nu\, {\langle{\dot n}_\nu\rangle\over C_I \langle \kappa_\nu\rangle},
\\[8pt]
\end{equation}
and only emitters within the volume defined by an absorption length contribute to the background intensity \citep{zuo93,madau99}. 
While, for a given ionizing photon emissivity, the local source approximation can overestimate the background intensity close to 
the hydrogen Lyman edge ($\nu\gta \nu_L$) at $z\sim 2$ (this is because a significant fraction of the emitted LyC photons 
at these epochs gets redshifted beyond the Lyman limit and does not contribute to the ionizing flux), it quickly approaches the exact value 
of $J_\nu$ at the high redshifts, $z\gta 5$, of interest here \citep{becker13}. In order to simplify our calculations, we shall adopt the source function approximation 
throughout the rest of this paper.

\section{Continuum Absorption}

The absorption opacity in the {post-reionization universe} is a crucial boundary condition for reionization models, and it is the highly ionized, 
post-reionization IGM that provides some of the most stringent empirical tests on the nature of cosmological ionizing sources. The technique of stacking quasar 
spectra provides a direct measurements of the photon proper mean free path at the hydrogen Lyman edge caused by the LLSs,
\begin{equation}
\langle \kappa^{\rm LLS}_{\nu_L}\rangle^{-1}=(37\pm 2)\,[(1+z)/5]^{-5.4\pm 0.4}\,{\rm Mpc}
\\[8pt]
\label{mfpLLS}
\end{equation}
in the interval $2.3<z<5.5$ \citep{worseck14,omeara13,prochaska09}, where the average is taken over all possible quasar sightlines.
The rapid evolution of this opacity exceeds that expected from cosmological expansion, 
indicating an increase in the number density and/or physical size of absorbing structures with redshift. In the limit of optically thick, Poisson-distributed 
clouds, and neglecting the cumulative photoelectric opacity of lower column density systems, the mean free path is 
equal to the average spacing between LLSs (``picket-fence" absorption), 
\begin{equation}
\langle \kappa^{\rm LLS}_{\nu}\rangle^{-1}={cH^{-1}\over (1+z)}\left({d{N}_{\rm LLS}\over dz}\right)^{-1}, 
\label{mfpdNdz}
\end{equation}
where $d{N}_{\rm LLS}/dz$ is the mean number of absorbers per unit redshift with hydrogen columns $N_\nHI>\sigma^{-1}_\nHI(\nu)$, $H(z)$ is the Hubble parameter, 
and $cH^{-1}/(1+z)=c|dt/dz|$ is the proper length interval in a Friedmann cosmology. Recent estimates of the mean free path based on the incidence of LLSs along 
the line of sight are in reasonable agreement with Equation (\ref{mfpLLS}) \citep{faucher08,songaila10,haardt12,rudie13,prochaska14}. 

To proceed further, we need a model of intergalactic absorption {\it before} the epoch of complete reionization. 
Our formalism assumes that the early IGM is organized in three main phases: a uniform ionized medium, a uniform neutral medium, and the LLSs,  which contain 
only a small fraction of the cosmic baryons. UV photons, with their short mean free path, carve out cosmological \HII\ regions in the uniform component
that expand in an otherwise largely neutral phase.\footnote{Such a description may not be accurate in the presence of hard-spectrum sources 
radiating copious penetrating X-ray photons. In the case of a galaxy-dominated reionization scenario, however, X-rays from compact binaries are expected to 
keep the electron fraction of the IGM between \HII\ cavities below 1 percent \citep{madau17}.}\ Let us denote, as is customary, the volume filling factor of such 
cosmological \HII\ regions as $Q$. Within the volume fraction $Q$, only the gas belonging to LLSs remains neutral because of its high density, whereas all the gas
is neutral within the fractional volume $(1-Q)$, i.e. $\langle x_\nHI\rangle=(1-Q)$. The volume-averaged neutral hydrogen density in the uniform IGM is 
$\langle n_\nHI\rangle=\langle x_\nHI\rangle \langle n_\nH\rangle$, and the corresponding absorption coefficient for photons of frequency $\nu_L<\nu<4\nu_L$ 
is 
\begin{equation}
\langle \kappa^{\rm IGM}_\nu\rangle=\langle n_\nH\rangle (1-Q)\sigma_\nHI(\nu). 
\label{mfpIGM}
\end{equation}
This quantity decreases rapidly during reionization as $Q\rightarrow 1$ and ionized bubbles merge and overlap. Throughout the volume, however, Lyman-limit absorbers 
in the outskirt of galaxies will still be consuming UV photons. The presence of these dense neutral clumps will not affect the topology of \HII\ regions as long 
as their mean spacing exceeds the typical ionized bubble radius. Once \HII\ regions grow beyond the mean separation between these systems, however, it is the LLSs 
rather than diffuse gas that determine the mean free path. Under the premises that Equation (\ref{mfpLLS}) correctly represents the volume-averaged opacity of LLSs  
and can be extrapolated to higher redshifts without significant loss in accuracy (since LLSs only affect the late stages of reionization), we can now estimate the volume-averaged, 
total absorption probability per unit length as 
\begin{equation}
\langle \kappa_\nu\rangle = \langle \kappa^{\rm IGM}_\nu\rangle +\langle \kappa^{\rm LLS}_\nu\rangle. 
\label{totmfp}
\end{equation}
Additional intuition on this result can be obtained by writing the effective optical depth per unit redshift interval of a clumpy IGM as 
\begin{equation}
{d\bar\tau\over dz}(\nu)=\int_0^{\infty}\, dN_\nHI\, {\partial^2 N\over \partial N_\nHI\partial z} (1-e^{-\tau}), 
\label{tauC}
\end{equation}
where $\partial^2 N/\partial N_\nHI\partial z$ is the joint frequency distribution in redshift and column density of Poisson-distributed absorbers 
along the line of sight, and $\tau=N_\nHI\sigma_\nHI(\nu)$ is the LyC optical depth through an individual absorber. 
Expanding the exponential in Equation (\ref{tauC}) to first order when $\tau(\nu)<1$ and neglecting it compared to unity when $\tau(\nu)>1$ gives 
\begin{equation}
\begin{split}
{d\bar \tau\over dz}(\nu) = & \, \sigma_\nHI(\nu)\int_0^{\sigma^{-1}_\nHI(\nu)}\, N_\nHI dN_\nHI\, {\partial^2 N\over \partial N_\nHI\partial z} \\
& + (dN_{\rm LLS}/dz).
\end{split}
\end{equation}
In the limit in which the LLSs are embedded in a uniform diffuse IGM, the first term on the right-hand side can be written as $\sigma_\nHI(\nu)\langle x_\nHI\rangle 
\langle n_\nH \rangle cH^{-1}/(1+z)$. Using Equations (\ref{mfpdNdz}) and (\ref{mfpIGM}), and writing the total average opacity as $\langle\kappa_\nu\rangle\equiv 
(d\bar\tau/dz)|dz/cdt|$, one then recovers Equation (\ref{totmfp}). 

An expression for the frequency-dependent opacity of LLSs can be derived by noting that the column density distribution of intervening 
absorbers can be fit with a set of power-laws, $dN/dN_\nHI\propto N_\nHI^{-\beta}$, with breaks at pivot points. Around neutral hydrogen columns of $N_\nHI\sim 
10^{17}\,$ cm$^{-2}$, uncertainties are still large, with recently estimated slopes at $\langle z\rangle=2.4$ ranging from $\beta\simeq 1.48$ \citep{rudie13} 
to $\beta\simeq 2.11$ \citep{prochaska14}. At frequencies $\nu>\nu_L$, the absorption coefficient can be shown to decrease as \citep[e.g.,][]{madau99}
\begin{equation}
\langle \kappa^{\rm LLS}_\nu\rangle= \langle \kappa^{\rm LLS}_{\nu_L}\rangle(\nu/\nu_L)^{3-3\beta},
\end{equation}
where we have approximated the photoionization cross section as $\sigma_\nHI\propto (\nu/\nu_L)^{-3}$.  
When $\beta=2$, this scaling matches that expected in a uniform absorbing medium, $\langle \kappa^{\rm IGM}_\nu\rangle\propto \sigma_\nHI(\nu)\propto (\nu/\nu_L)^{-3}$. 
While we shall adopt $\beta=2$ in all our calculations below, in practice our results are rather insensitive to this choice because of the steep frequency 
dependence of the photoelectric cross section and galaxy UV spectrum, which also enter in the hydrogen photoionization rate. 

We shall use this simple, semi-empirical approach to clumpy IGM absorption to model photon sink processes on large scales during reionization. A different 
treatment was presented by \citet{miralda00}, who used a volume-weighted density distribution of IGM gas, $P_V(\Delta_b)$ (where $\Delta_b=\rho_b/\bar \rho_b$), 
to describe the ionization state of an inhomogeneous  universe. They assumed that reonization proceeds ``outside-in", first into the underdense voids and then more 
gradually into overdense regions, and that at any given epoch all the gas with density above a critical overdensity $\Delta_i$ remains neutral, self-shielded from the 
ionizing background, while all lower density material is completely ionized in a fraction $Q$ of the volume. Additionally, their model postulates that the absorption 
mean free path scales as $F_V(\Delta_i)^{-2/3}$, where $F_V(\Delta_i)$ is the volume filling factor of gas with $\Delta_b>\Delta_i$. The main advantage of this technique 
is that it allows one to calculate the mean free path and the ``recombination clumping factor" based on a realistic density distribution. The main disadvantage is that
numerical simulations appear to support the opposing, ``inside-out" view, where reionization proceeds from high- to low-density regions, and the gas density and fractional
ionization are correlated on large scales \citep[e.g.,][]{ciardi03,iliev06,finlator09,friedrich11,so14,bauer15}. In our model, high-density regions (the 
LLSs) control the photon mean free path only in the very late stages of reonization and after overlap, while the low-density neutral IGM 
provides the bulk of the opacity in the early stages. To improve further on our treatment, we will account for inhomogeneities in 
the ionized diffuse phase via an effective recombination timescale derived from cosmological hydrodynamics simulations. 

\section{Reionization Equation}

\subsection{Hydrogen Photoionization Rate}

To make headway towards a reionization equation that can be solved analytically, we start by writing the volume-averaged hydrogen photoionization rate 
in the local approximation as
\begin{equation} 
\begin{split}
\langle \Gamma_\nHI\rangle = &\int d\nu\, {\langle {\dot n}_\nu\rangle\,\sigma_\nHI(\nu)\over \bar\kappa_\nu}\\
= & \int d\nu\, {\langle {\dot n}_\nu\rangle\,\sigma_\nHI(\nu)\over C_I(\langle \kappa^{\rm IGM}_\nu\rangle + \langle \kappa^{\rm LLS}_\nu\rangle)}.
\label{eq:Gamma}
\end{split}
\end{equation}
The photon spectrum of star-forming galaxies between 1 and 4 Ryd can be approximated by a power-law, ${\dot n}_\nu\propto \nu^{-2}$ \citep[e.g.,][]{kewley01}. 
Ignoring spatial and temporal variations in the spectral shape of the ionizing photon emissivity, evaluating the integral in Equation (\ref{eq:Gamma}) 
analytically for $\beta=2$, and using the relation \citep[e.g.,][]{kohler07,onorbe17} 
\begin{equation} 
\langle x_\nHI \Gamma_\nHI\rangle=C_I \langle x_\nHI \rangle\ \langle \Gamma_\nHI \rangle, 
\end{equation}
we finally obtain an expression for the volume-averaged photoionization rate per hydrogen atom
\begin{equation} 
\langle x_\nHI \Gamma_\nHI\rangle={\langle {\dot n}_{\rm ion}\rangle \over \langle n_\nH\rangle}\,{1\over (1+\langle \kappa^{\rm LLS}_{\nu_L}\rangle/
\langle \kappa^{\rm IGM}_{\nu_L}\rangle)}, 
\label{eq:photo}
\end{equation}
where $\langle {\dot n}_{\rm ion} \rangle=\int_{\nu_L} d\nu \langle {\dot n}_\nu\rangle$.
{\it Note that the volume-averaged photoionization rate is independent of the photoionization clumping factor $C_I$, and can therefore be computed without 
any knowledge of the cross-correlation between the fractional abundance of \HI\ and the radiation field.} 
 
\subsection{Radiative Recombination Rate}

Let us now recast the hydrogen recombination term in Equation (\ref{eq:xHII}) as
\begin{equation} 
\langle n_e x_\nHII \alpha_\nHII(T)\rangle \equiv {\langle x_\nHII\rangle\over \bar t_{\rm rec}}={Q\over \bar t_{\rm rec}}, 
\\[8pt]
\label{eq:recom}
\end{equation}
where $\bar t_{\rm rec}$ is the effective recombination timescale of a clumpy IGM.\footnote{This is not the volume-averaged recombination time, 
which is defined as $\langle t_{\rm rec}\rangle=\langle [n_e \alpha_\nHII(T)]^{-1}\rangle$ and weighs preferentially low-density regions.}\, 
Because it scales quadratically with density, the volume-averaged recombination rate depends on the actual gas density distribution within the volume.   
It is often convenient to rewrite $\bar t_{\rm rec}$ as \citep{madau99}
\begin{equation}
\bar t_{\rm rec,M}={1\over (1+\chi)\langle n_\nH \rangle\alpha(T_0) C_R},
\label{eq:tmadau}
\end{equation}
where $\chi=Y/4X=0.083$ accounts for the presence of photoelectrons from \HeII, $\alpha$ is the recombination coefficient at temperature $T_0$, 
and $C_R$ is the recombination clumping factor that is evaluated with the help of numerical simulations of reionization (which may or may not satisfy all existing 
observational constraints). In the above expression, $C_R$ corrects for the enhanced 
recombination rate induced by structure formation relative to a highly ionized IGM of uniform density and uniform and constant temperature at all times. 
A number of different definitions of the clumping factor exist in the literature, with different cuts in baryon overdensities, ionization and metallicity 
levels, and including (or not) the temperature dependence of the recombination coefficient \citep[see, e.g.][]{kohler07,pawlik09,finlator12,shull12,jeeson14,kaurov14,so14}. 
Values of $C_R$ of order a few at $z\gta 6$ are typically derived in recent work if the clumping factor is averaged only over gas with
overdensity $\Delta_b<100$ (a threshold that excludes dense gas bound to halos as is done for the photoionization clumping factor $C_I$), but there are uncertanties.\footnote{Photon 
losses by LLSs are also due to radiative recombinations. In analytical models it is standard practice, however, to include the effect of LLSs as a reduction in 
the source term through the finite mean free path of ionizing radiation. Three different quantities -- the escape fraction, the clumping factor, and the mean free 
path -- are then used to describe what are essentially radiative recombinations in the ISM, the IGM, and the LLSs. Numerical simulations of cosmic reionization 
show that such a separation into distinct regimes may indeed be reasonable \citep{kaurov15}.}\,
In the following, we shall insert into $\bar t_{\rm rec,M}$ the expression $C_R=2.9[(1+z)/6]^{-1.1}$ from \citet{shull12}, which gives a 
clumping factor similar to the one derived by \citet{finlator12}. We shall also use the Case-A recombination coefficient, $\alpha_A$,  
as recombination photons are typically redshifted below threshold before being absorbed \citep{kaurov14}. The temperature of ionized 
gas will be fixed to $T_0=10^{4.3}\,$K to account for the photoheating associated with the reionization process itself \citep{hui03}.
Note that gas at this temperature recombines in Case-A at the same rate of gas in the oft-used Case-B situation at $T_0=10^{4}\,$K. 
Average temperatures higher than $10^4\,$K are found for  the recombining IGM at $5<z<10$ in simulations \citep{so14}.

A fitting formula to $\bar t_{\rm rec}$ has been recently provided by \citet{so14} that involves no ad hoc clumping factor, 
\begin{equation} 
\bar t_{\rm rec,S}=2.3\,[(1+z)/6]^{-4.35}\,{\rm Gyr}, 
\label{eq:trec}
\end{equation}
and is based on the analysis of a fully coupled radiation hydrodynamical realization of hydrogen reionization that begins at $z\simeq 10$ and completes at $z\simeq 5.8$. 
This is the actual, appropriately-averaged recombination timescale in the simulation, again after applying a gas overdensity threshold of $\Delta_b<100$.
The recombination time $\bar t_{\rm rec,M}$ is a factor 1.6 shorter than $\bar t_{\rm rec,S}$ at $z=6$, and 1.9 times longer 
at $z=10$. The two timescales are equal at $z=7.5$, and are both shorter than the Hubble time at all redshifts $z\gta 6$. 
In an attempt to bracket the uncertainty in the recombination timescale of the IGM, we shall use both $\bar t_{\rm rec,M}$ and $\bar t_{\rm rec,S}$ 
in our numerical estimates below.

\begin{figure*}
\centering
\includegraphics*[width=0.49\textwidth]{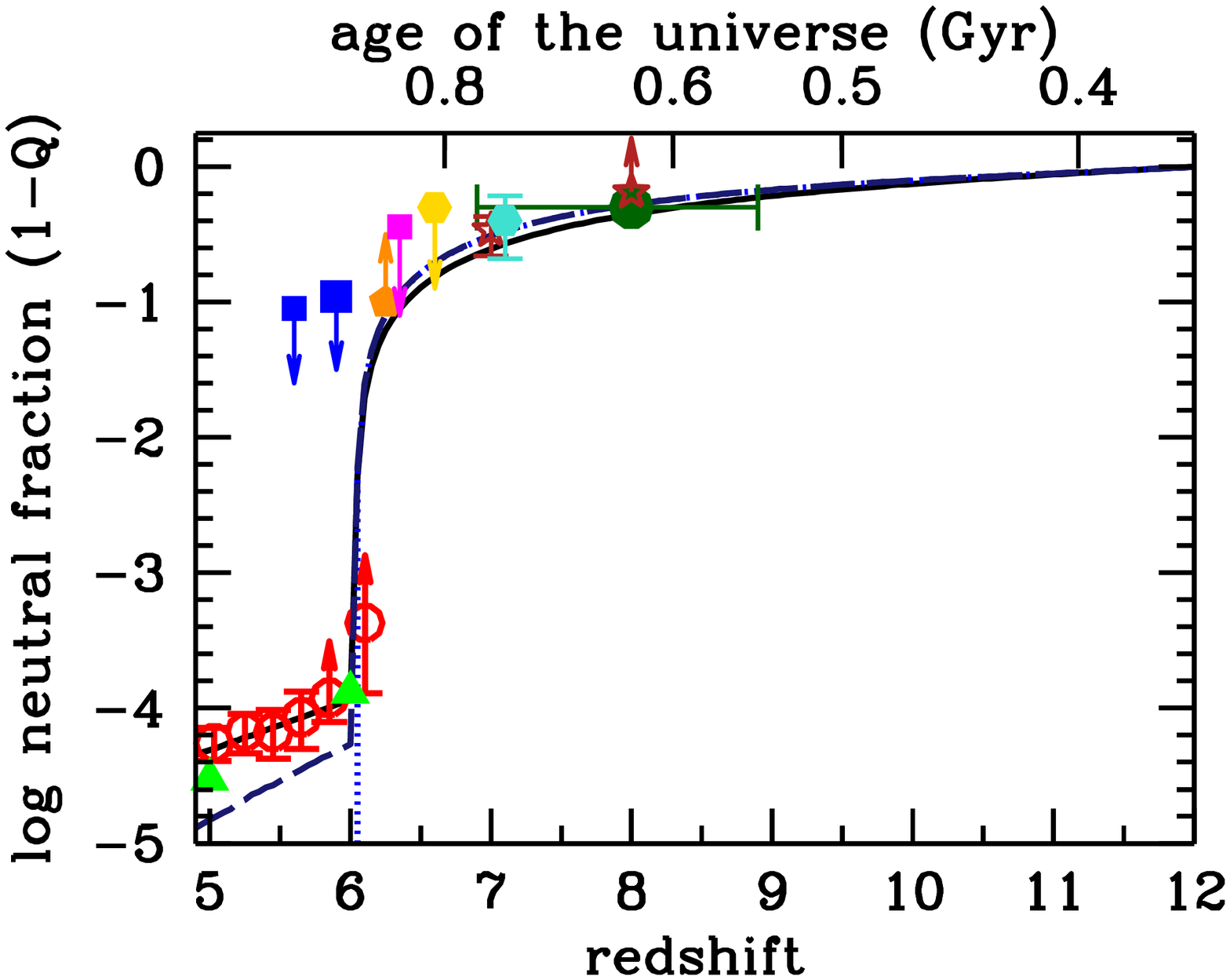}
\includegraphics*[width=0.49\textwidth]{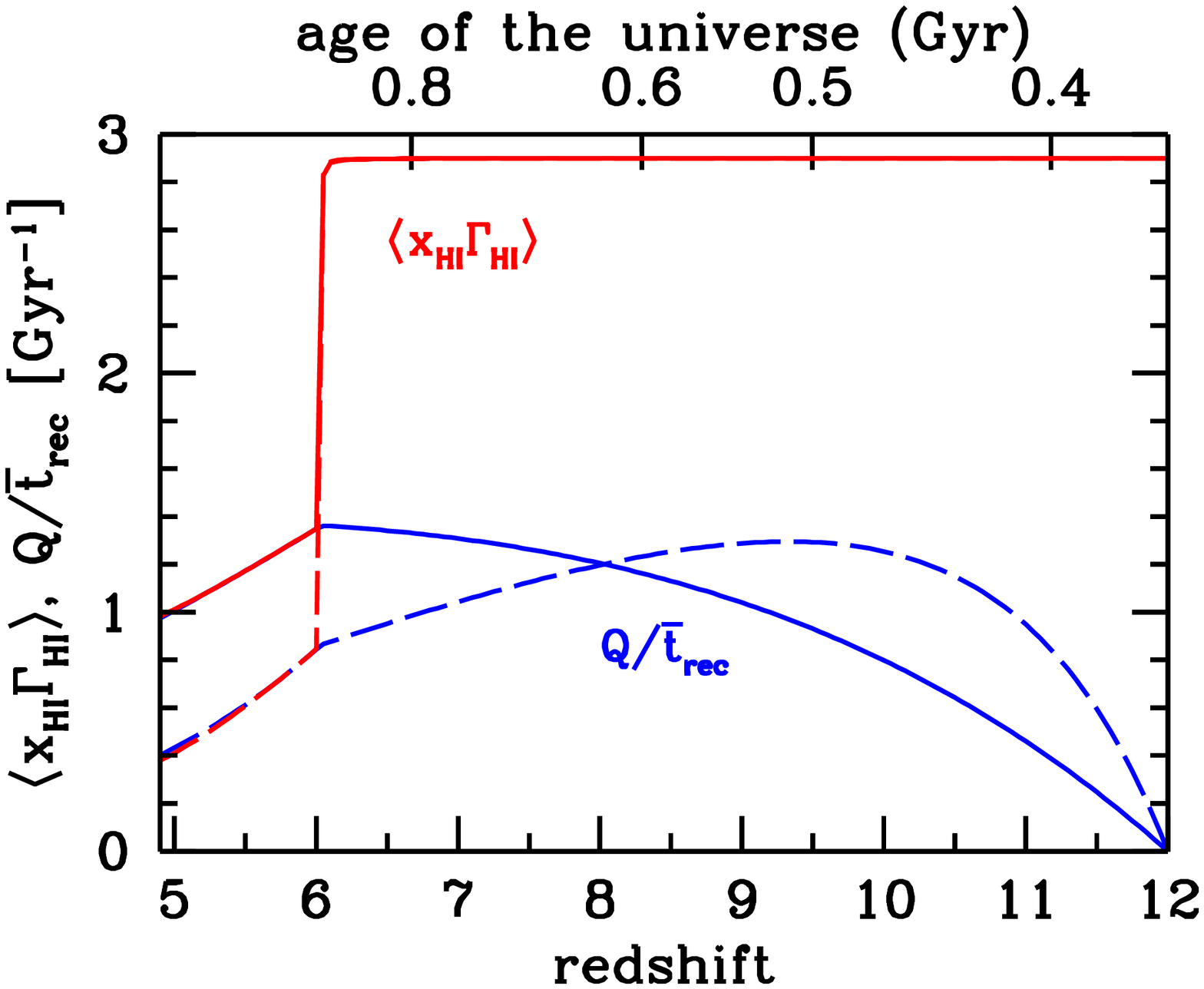}
\caption{Reionization histories predicted by integrating Equation (\ref{eq:dQdt}) with a constant emission rate of ionizing photons per hydrogen 
atom, $\langle \dot n_{\rm ion}\rangle/\langle n_\nH \rangle=2.9$ Gyr$^{-1}$.
Left panel: Average neutral fraction $(1-Q)$ of an IGM that recombines according to Equation (\ref{eq:tmadau}) (solid curve) or Equation (\ref{eq:trec}) 
(dashed curve). In both models the photon mean free path includes a contribution from the LLSs.  
The dotted line shows a model without LLSs and with $\bar t_{\rm rec}=\bar t_{\rm rec,S}$. The curves illustrate that the process of reionization is 
quite extended: even if it completes at redshift 6, the typical location in the IGM is 50\% likely to be fully ionized already at $z\sim 8$. 
The data points represent constraints on the ionization state of the IGM from: the Gunn-Peterson optical depth at $5<z<6.1$ (red open circles, 
\citealt{fan06}), measurements of the \Lya\ forest opacity combined with hydrodynamical simulations (green triangles, \citealt{bolton07}),
the dark pixel statistics at $z=5.6$ and $z=5.9$ (blue squares, \citealt{mcgreer15}),
the gap/peak statistics at $z=6.32$ (magenta square, \citealt{gallerani08}),
the damping wing absorption profiles in the spectra of quasars at $z=6.3$ (orange pentagon, \citealt{schroeder13}) 
and $z=7.1$ (turquoise hexagon, \citealt{greig17,mortlock11}), the redshift-dependent prevalence of LAEs in narrow-band surveys at $z=7$ and $z=8$
(firebrick stars, \citealt{schenker14}) and their clustering properties at $z=6.6$ (gold hexagon, \citealt{ouchi10}).
The recent limit on the redshift at which $\langle x_\nHI\rangle=0.5$,
extracted from the {\it Planck} CMB data \citep{planck16}, is plotted as the green dot.
Right panel: Relative contribution of the photoionization source term, $\langle x_\nHI\Gamma_\nHI\rangle$ (red lines), and sink recombination term, 
$Q/\bar t_{\rm rec}$ (blue lines), in the reionization equation. The solid and dashed curves correspond to the different expressions 
for the recombination timescale $\bar t_{\rm rec,M}$ and $\bar t_{\rm rec,S}$, respectively.
Photoionizations dominate over recombinations until just before overlap; the two terms come swiftly into balance at overlap, and remain in equilibrium thereafter.
}
\label{fig1}
\end{figure*}

\subsection{Reionization Revisited}

Substituting Equations (\ref{eq:photo}) and (\ref{eq:recom}) into the ionization balance equation (\ref{eq:xHII}) for hydrogen, we finally obtain
\begin{equation}
{dQ\over dt} = {\langle \dot n_{\rm ion}\rangle \over \langle n_\nH \rangle (1+\langle \kappa^{\rm LLS}_{\nu_L}\rangle/\langle \kappa^{\rm IGM}_{\nu_L}\rangle)}  
-{Q\over \bar t_{\rm rec}}. 
\label{eq:dQdt}
\end{equation}
{\it This is our revised reionization equation}. Prior to overlap, when $Q<1$ and $\langle \kappa^{\rm LLS}_{\nu_L}\rangle\ll \langle \kappa^{\rm IGM}_{\nu_L}\rangle$, 
the source term is 
\begin{equation}
\langle x_\nHI \Gamma_\nHI \rangle\sim {\langle {\dot n}_{\rm ion}\rangle\over \langle n_\nH\rangle},
\end{equation}
i.e. the volume-averaged photoionization rate becomes independent of the mean free path and equal to the ionizing photon emission rate into the IGM per hydrogen atom. 
In this limit, we recover the standard reionization Equation (\ref{eq:intro}).
After overlap, as $Q\rightarrow 1$ and $\langle \kappa^{\rm LLS}_{\nu_L}\rangle\gg \langle \kappa^{\rm IGM}_{\nu_L}\rangle$, the source term becomes
\begin{equation}
\langle x_\nHI \Gamma_\nHI \rangle\sim {\langle {\dot n}_{\rm ion}\rangle\over \langle n_\nH\rangle}
{\langle \kappa^{\rm IGM}_{\nu_L}\rangle \over \langle \kappa^{\rm LLS}_{\nu_L}\rangle} \propto (1-Q),
\end{equation}
i.e.  the volume-averaged photoionization rate is reduced by the finite mean free path of LLSs and becomes smaller as the neutral fraction of the 
IGM decreases, as expected on physical grounds. This new formulation explicitly accounts for the presence of optically thick absorbers that cause the mean 
free path of LyC photons to remain small even after overlap, and $Q$ never exceeds unity. {The integration of Equation (\ref{eq:dQdt}) therefore 
provides a link between the pre-overlap and post-overlap phases of the reionization process.}

\section{The IGM as a Photon Counting Device}

We start by considering a simple model for the global ionization history of the universe, with the goal of providing an illustrative description of 
how reionization may proceed over cosmic time following Equation (\ref{eq:dQdt}).  Figure \ref{fig1} shows theoretical curves for $(1-Q)$ and $dQ/dt$ 
obtained by numerically integrating Equation (\ref{eq:dQdt}) from $z=12$ onwards, assuming a constant emission rate of ionizing 
photons per hydrogen atom, $\langle \dot n_{\rm ion}\rangle/\langle n_\nH \rangle=2.9$ Gyr$^{-1}$. This value corresponds to a comoving ionizing photon emissivity 
into the IGM of 
\begin{equation} 
{\langle \dot N_{\rm ion}\rangle} \equiv \langle \dot n_{\rm ion}\rangle/(1+z)^3=5.1\times 10^{50}\,{\rm s^{-1}\,Mpc^{-3}}.
\label{eq:cemiss1}
\end{equation}
\citet{becker13} used measurements of the mean \Lya\ and LyC opacity to estimate the ionizing emissivity in the post-reionization era, and found that this is 
in the range $\langle \dot n_{\rm ion}\rangle/\langle n_\nH \rangle=$2--14 Gyr$^{-1}$ at redshift $4.75$. The fiducial constant value adopted here at $5<z<12$ is at the
low end of this range, and leads to a reionization history (left panel) that is consistent with a number of observational constraints on the ionization 
state of the $z>5$ universe, from the redshift-dependent prevalence of LAEs in narrow band surveys at $z=$7--8 \citep{schenker14}, 
to the damping wing absorption profiles measured in the spectra of $z=6.3$ \citep{schroeder13} and $z=7.1$ quasars \citep{greig17,mortlock11}.
{The redshift of overlap corresponds to an integrated output into the IGM of about 2.5 ionizing photons per hydrogen atom.} 
The redshift-asymmetric parameterization for the evolution of the ionized fraction adopted in \citet{planck16},
\begin{equation} 
Q=\left({z_{\rm early}-z\over z_{\rm early}-z_{\rm end}}\right)^\alpha,
\label{eq:rspara}
\end{equation}
where $z_{\rm early}=20$ is the redshift around which the first emitting sources form and $z_{\rm end}$ is the redshift at which reionization ``ends" ($Q=0.99$), 
provides a good fit to the reionization histories in the left panel of Figure \ref{fig1} for $10>z>z_{\rm end}=6$ and exponent $\alpha=4$. 
Denoting with $z_{20}$ the redshift at which $Q=0.2$, our models yield a duration of reionization -- the pre-overlap phase of \citet{gnedin00} --
in the range $z_{20}-z_{\rm end}=4.0-4.3$, which is compatible with constraints on patchy reionization based on the kinetic Sunyaev-Zeldovich (kSZ) effect, 
$z_{20}-z_{\rm end}<5.4$ (95\% CL, \citealt{george15}). Also plotted in the figure is the recent limit on the redshift of reionization $z_{50}$ (the redshift 
at which the hydrogen ionized fraction is 50\%) extracted from the {\it Planck} CMB data \citep{planck16}: $z_{50}=8.0^{+0.9}_{-1.1}$ (uniform prior,
redshift-asymmetric parameterization).

The impact of the LLSs in shaping the end of the reionization process and keeping the mean hydrogen neutral fraction of the IGM above $10^{-5}$ after overlap 
is clearly visible, as well as the sensitivity of the ionization state at these epochs to the recombination timescale. In this toy model, an IGM that recombines with the 
timescale $\bar t_{\rm rec,M}$ provides a better fit to observations of the Gunn-Peterson optical depth in the spectra of 
Sloan Digital Sky Survey (SDSS) quasars at $5<z<6.1$ \citep{fan06}, as well as to estimates of
the ionized fraction at $z=5$ and $6$ based on a combination of hydrodynamical simulations with
measurements of the \Lya\ forest opacity \citep{bolton07}.
While the pre-overlap stages extend over a considerable range of redshift, the phase of overlap, indicated by the sudden drop in the neutral gas fraction at $z\simeq 5.9$, 
is clearly defined even in the presence of the LLSs: $\langle x_\nHI\rangle$ decreases by more than 2 orders of magnitude over a fraction of the then Hubble time.
Such a dramatic transition marks the epoch when the photon mean free path becomes determined by the LLSs rather than by the typical size of \HII\ regions. 
Note how, in practice, the source term in our reionization equation becomes {\it independent} of the actual size of \HII\ bubbles in the pre-overlap stages 
provided this is much smaller than the spacing between LLSs, i.e. in the limit $\langle \kappa^{\rm IGM}_{\nu_L}\rangle \gg \langle 
\kappa^{\rm LLS}_{\nu_L}\rangle$.

The right panel in Figure \ref{fig1} shows the relative contribution of the source and sink terms to Equation (\ref{eq:dQdt}). 
Photoionizations dominate over recombinations until just before overlap. As overlap is approached, the neutral fraction of the IGM dives below 
$10^{-4}$ and the photoionization source term drops rapidly. Photoionizations come into balance with recombinations for the first time at overlap, 
where $dQ/dt\rightarrow 0$ and
\begin{equation}
\langle \dot n_{\rm ion}\rangle \bar t_{\rm rec}=\langle n_\nH \rangle (1+\langle \kappa^{\rm LLS}_{\nu_L}\rangle / \langle \kappa^{\rm IGM}_{\nu_L}\rangle), 
\\[8pt]
\label{eq:Q1}
\end{equation}
and remain so thereafter. This relation modifies the original criterion by \citet{madau99}, 
\begin{equation}
\langle \dot n_{\rm ion}\rangle \bar t_{\rm rec}=\langle n_\nH \rangle, 
\\[8pt]
\label{eq:oldQ1}
\end{equation}
for the ionizing photon emissivity necessary to maintain a clumpy, recombining IGM in a highly ionized state in the absence of LLSs. 
Just before overlap, however, the factor $(1+\langle \kappa^{\rm LLS}_{\nu_L}\rangle / \langle \kappa^{\rm IGM}_{\nu_L}\rangle)$ is very close to unity 
and the number of ionizing photons emitted into the IGM in one recombination time is equal to the number of hydrogen atoms, as in the criterion by 
\citet{madau99}.\footnote{In our model, the factor ($1+\langle \kappa^{\rm LLS}_{\nu_L}\rangle/\langle \kappa^{\rm IGM}_{\nu_L}\rangle$)
grows rapidly after overlap. In the case $\bar t_{\rm rec}=\bar t_{\rm rec,M}$, for example, 
this factor is 1.007 at $z=6.1$ (neutral fraction 0.02), 2.15 at $z=6.0$ (neutral fraction $10^{-4}$), and 2.87 at $z=5$ 
(neutral fraction $5\times 10^{-5}$). The increase with decreasing redshift is even faster for $\bar t_{\rm rec}=\bar t_{\rm rec,S}$.
}\, 
%
%

It has been recently argued by \citet{so14} that the time-dependent differential Equation (\ref{eq:dQdt}) and its quasi-instantaneous
version, Equation (\ref{eq:Q1}), should not be used to predict the epoch of reionization completion.
This is because, according to \citet{so14}, Equation (\ref{eq:Q1}) ignores history-dependent terms in the global ionization balance that are not ignorable, and 
both equations systematically overestimate the redshift of reionization completion compared to their simulations because the conversion 
of LyC photons into new ionized hydrogen atoms becomes inefficient at late times. The validity of the quasi-instantaneous approximation 
as predictor of reionization completion/maintenance is, however, clearly shown in the right panel of Figure \ref{fig3}. We believe 
instead that the disagreement is an artifact of the simulations underestimating the amount of UV absorption in the relatively small box adopted.  
According to \citet{so14}, the ratio of photoionizations to emitted UV photons decreases as $Q\rightarrow 1$, and is as low as
5\% at overlap. By contrast, in our model, even as $Q\rightarrow 1$, the mean free path close to the Lyman edge is much smaller than the horizon because of the 
presence of LLSs, and the ``ionization efficiency" remains 100\%. It is the insufficient absorption of LyC radiation by the LLSs, unresolved in the 
\citet{so14} simulations, which causes the early flattening in the $Q(z)$ curve observed by \citet{so14} and not seen in our analytic calculations.

Together with evidence of patchy ionization in the IGM from an extreme \Lya\ Gunn-Peterson trough at redshift 5.8 \citep{becker15}, the observational results 
listed above strongly suggest that cosmic reionization was ``completed" (i.e. the universe became ionized at more than 99\%)
around redshift 6, and that above redshift 10 hydrogen was ionized to less than 10\% levels. In particular, the latest {\it Planck} analysis disfavour an 
early beginning of reionization and yields a Thomson scattering optical depth $\tau_{\rm es}=0.058\pm 0.012$ \citep{planck16}. The toy models in Figure \ref{fig1} 
are consistent by design with the {\it Planck} determination, with 
\begin{equation}
\tau_{\rm es}=c\sigma_T \langle n_\nH \rangle \int_0^{12}{(1+z')^2dz'\over H(z')}Q(1+\chi)\simeq 0.06,
\label{eq:taues}
\end{equation}
where $\sigma_T$ the Thomson cross section, and we have ignored the small contribution of photoelectrons from the double reionization of helium at late times.
Also note how, taken at face values, the constraints from the Gunn-Peterson opacity, the dark pixel and gap/peak statistics, and the damping wing absorption 
profiles in the spectra of quasars all appear to require a very sharp transition around redshift 6 from a $\sim 10$\% neutral to a nearly fully ionized universe. 

\begin{figure*}
\centering
\includegraphics*[width=0.49\textwidth]{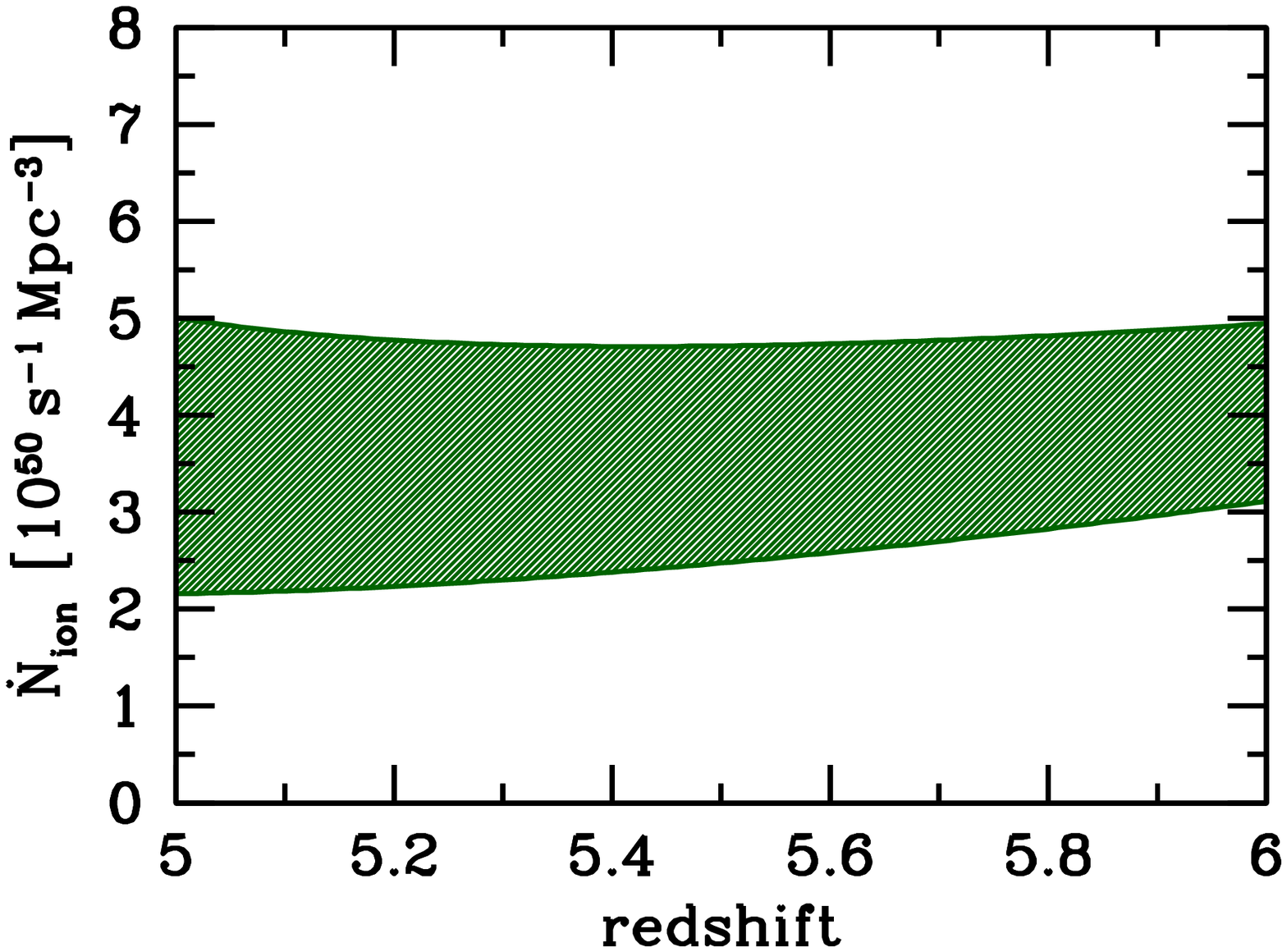}
\includegraphics*[width=0.49\textwidth]{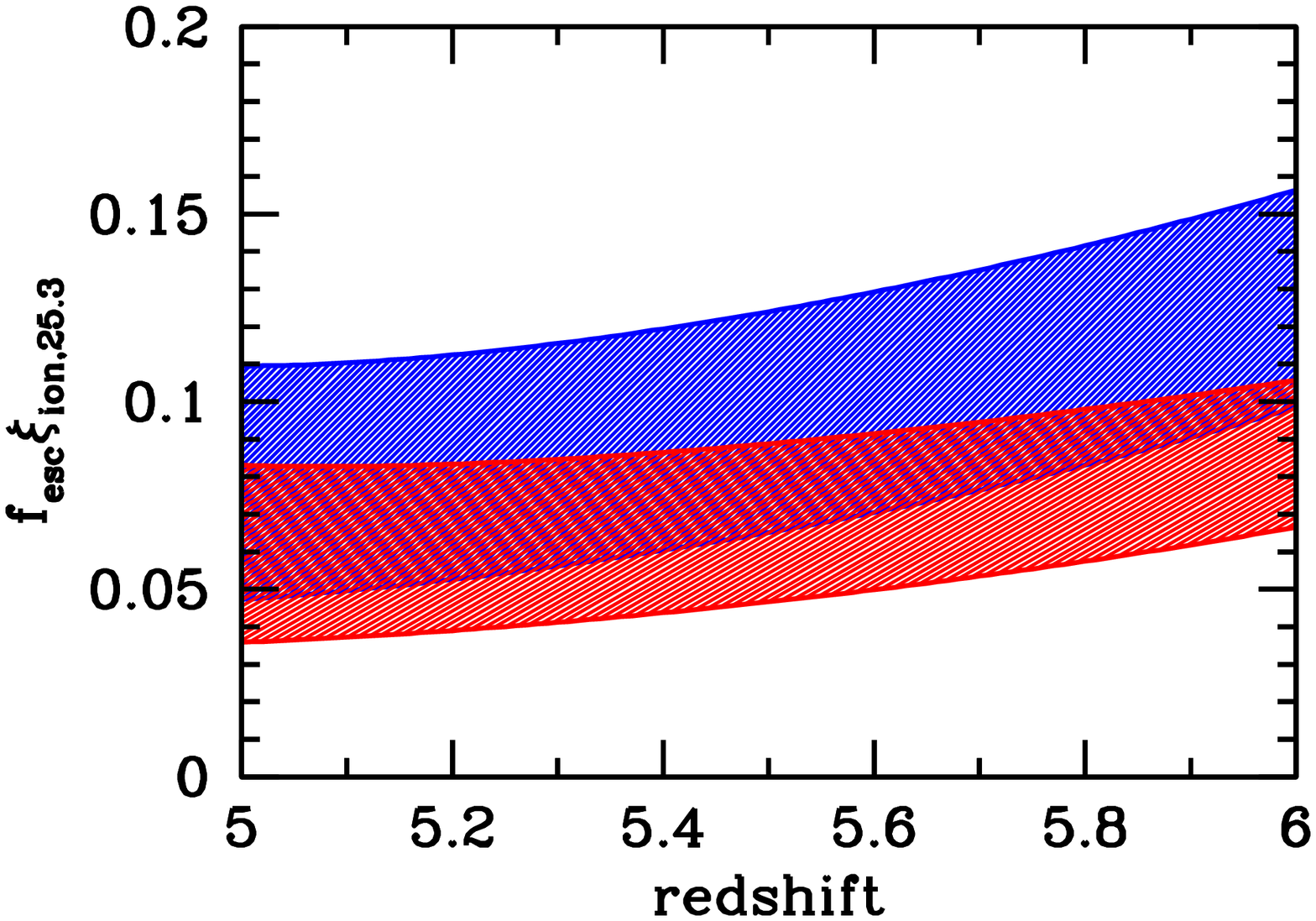}
\caption{Constraining the sources of photoionizations in the post-overlap era. 
Left panel: The comoving ionizing photon emissivity into the IGM that is required to reproduce the Gunn-Peterson optical depth measured at $5<z<6$ 
in the spectra of SDSS quasars \citep{fan06}. The shading highlights the impact of using different expressions for the IGM recombination timescale, 
$\bar t_{\rm rec}=\bar t_{\rm rec,M}$ (upper envelope), and $\bar t_{\rm rec}=\bar t_{\rm rec,S}$ (lower envelope). 
Right panel: Limits on the ``LyC photon return into the IGM" of star-forming galaxies, $f_{\rm esc}\xi_{\rm ion}$, set by the same data. 
The two models of the  galaxy comoving UV luminosity density given in Equations (\ref{eq:rho16}) and (\ref{eq:rho13}) are shown with the blue
and red bands, respectively. The quantity $\xi_{\rm ion,25.3}$ is expressed in units of $10^{25.3}\,\xiunits$. 
}
\label{fig2}
\end{figure*}

\section{Discussion}

It is widely believed that normal, star-forming galaxies dominate the production of LyC radiation in the pre-overlap, overlap, and immediate post-overlap stages of cosmic 
hydrogen reionization. The modeling in the previous sections offers an opportunity for a fresh look at this scenario.

\subsection{Post-Overlap}

In the post-overlap era, defined here as the epoch corresponding to $5<z<6$, observations of the Gunn-Peterson optical depth in the spectra of SDSS quasars 
can be fit with a volume-average neutral fraction $\langle x_\nHI\rangle\sim 7.2\times 10^{-5}(z-4.3)$ \citep{fan06}. 
Equations (\ref{mfpLLS}) and (\ref{mfpIGM}) imply then

\begin{equation}
{\langle \kappa^{\rm LLS}_{\nu_L}\rangle\over \langle \kappa^{\rm IGM}_{\nu_L}\rangle}=1.2\times 10^{-6}{(1+z)^{2.4}\over \langle x_\nHI\rangle}
\sim 0.017{(1+z)^{2.4}\over (z-4.3)}.
\end{equation}
Inserting this expression into Equation (\ref{eq:Q1}), we derive from the photoionization balance condition after overlap the required 
comoving ionizing photon emissivity into the IGM,
\begin{equation}
\begin{split}
{\langle \dot N_{\rm ion}\rangle} & ={\langle n_{\nH} \rangle\over (1+z)^3\bar t_{\rm rec}}
(1+\langle \kappa^{\rm LLS}_{\nu_L}\rangle / \langle \kappa^{\rm IGM}_{\nu_L}\rangle)\\
& ={2-5}\times 10^{50}\,{\rm s^{-1}\,Mpc^{-3}},
\label{eq:cemiss2}
\end{split}
\end{equation}
where the range of numerical values covers the redshift interval $5<z<6$ and the uncertanties in the recombination timescale (see the left panel of Fig. \ref{fig2}).

Over the redshift range $4<z<10$, the galaxy UV (1500\,\AA) comoving luminosity density, $\rho_{\rm UV}$, obtained by integrating the observed LF down 
to the absolute magnitude cut-off $M_{\rm lim}$, evolves approximately as 
\begin{equation}
\log \rho_{\rm UV}\,(\lumdens)=26.20-0.16(z-6) 
\label{eq:rho16}
\end{equation}
(``low $\rho_{\rm UV}$ model") for $M_{\rm lim}=-16$ mag, and as 
\begin{equation}
\log \rho_{\rm UV}\,(\lumdens)=26.37-0.11(z-6) 
\label{eq:rho13}
\end{equation}
(``high $\rho_{\rm UV}$ model") when the integration is carried down to $M_{\rm lim}=-13$ \citep{bouwens15planck}. 
A turn-over in the $z\sim 7$ LF at these faint luminosities is suggested by combining abundance matching with detailed studies of the
color-magnitude diagram of low-luminosity dwarfs in the Local Group \citep{boylan15}. According to stellar population synthesis models, the LyC photon yield per 
unit UV luminosity at 1500\,\AA\ is $\log \xi_{\rm ion}/[\xiunits]=25.2-25.3$ for a metal-poor stellar population with no dust 
\citep[e.g.,][]{schaerer03,robertson13,madau14}.\footnote{Note that, in 
this context, the symbol $\xi_{\rm ion}$ is sometimes used to express the LyC photon yield per unit rate of star formation instead \citep[e.g.][]{gnedin16}.}\, 
These values are consistent with those derived in a sample of sub-$L^*$ galaxies at $z=$4--5 by \citet{bouwens16Lyc} and based on IRAC-inferred H$\alpha$ measurements.
We can now write the comoving photon emission rate into the IGM as $\langle \dot N_{\rm ion}\rangle = f_{\rm esc}\xi_{\rm ion}\rho_{\rm UV}$,
where $f_{\rm esc}$ is the globally-averaged relative fraction of LyC photons that escape from individual galaxies and make it into the IGM. 
The condition for star-forming galaxies to be the source of the cosmic ionizing emissivity {\it after overlap} sets then a constraint on the 
product of the two unknowns $f_{\rm esc}\xi_{\rm ion}$, the ``ionizing photon return into the IGM", 
\begin{equation}
f_{\rm esc}\xi_{\rm ion} = {\langle n_{\nH} \rangle (1+\langle \kappa^{\rm LLS}_{\nu_L}\rangle / \langle \kappa^{\rm IGM}_{\nu_L}\rangle)
\over \rho_{\rm UV}(1+z)^3\bar t_{\rm rec}}.
\\[8pt]
\end{equation}
This is shown in the right panel of Figure \ref{fig2}, where the quantity $\xi_{\rm ion,25.3}$ is expressed in units of $10^{25.3}\,\xiunits$,
The shadings reflect the uncertanties in the recombination timescale, while the upper and lower swaths 
correspond to magnitude cut-offs of $M_{\rm lim}=-16$ mag and $M_{\rm lim}=-13$ mag in the luminosity density, respectively.  
At these epochs, values of $f_{\rm esc}\xi_{\rm ion,25.3}\simeq$ 0.04--0.11 can match the data at redshift 5, with a 
trend toward increasing ionizing photon returns with increasing redshifts. As shown below, the same trend may also hold true in the pre-overlap era.

\begin{figure*}
\centering
\includegraphics*[width=0.49\textwidth]{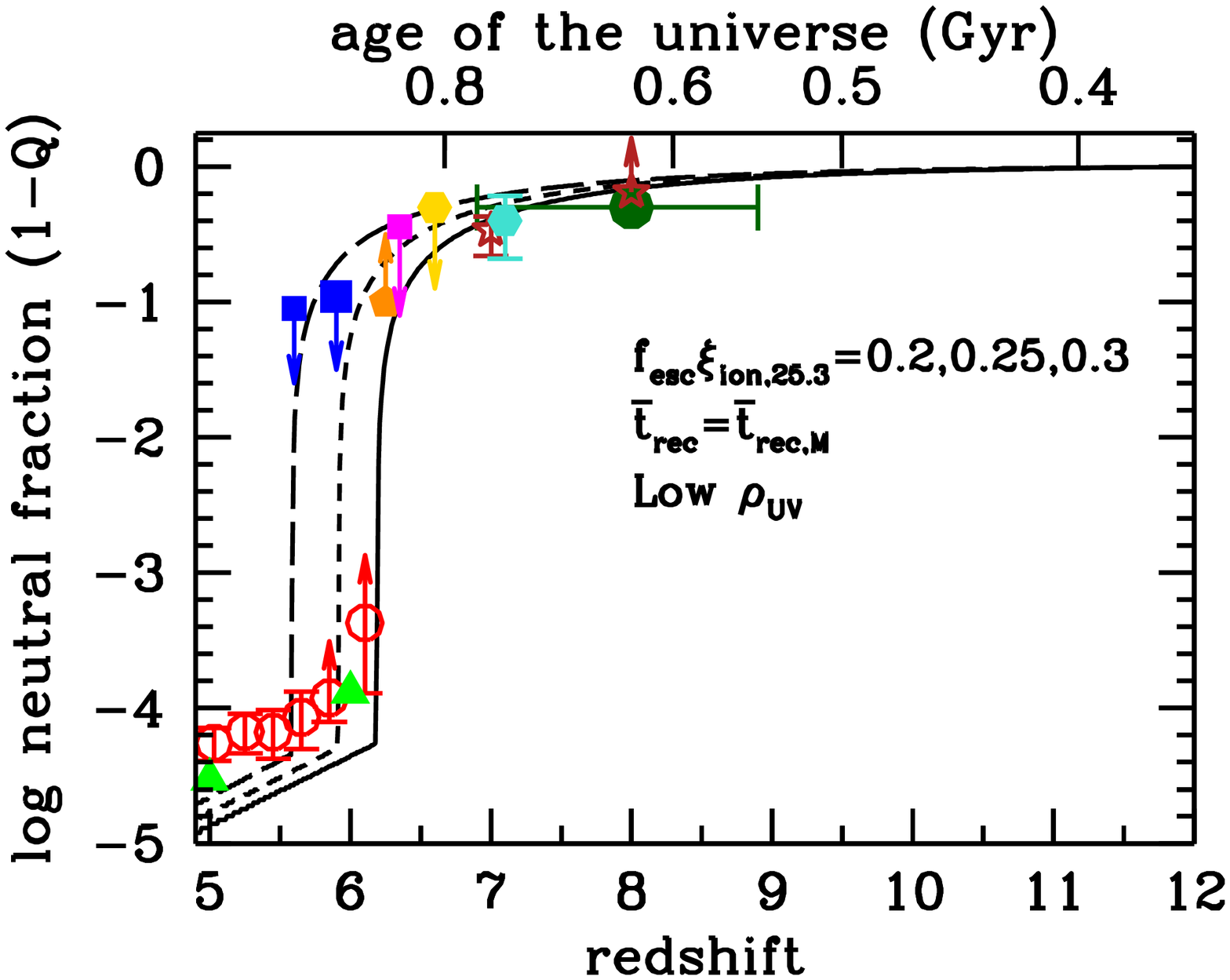}
\includegraphics*[width=0.49\textwidth]{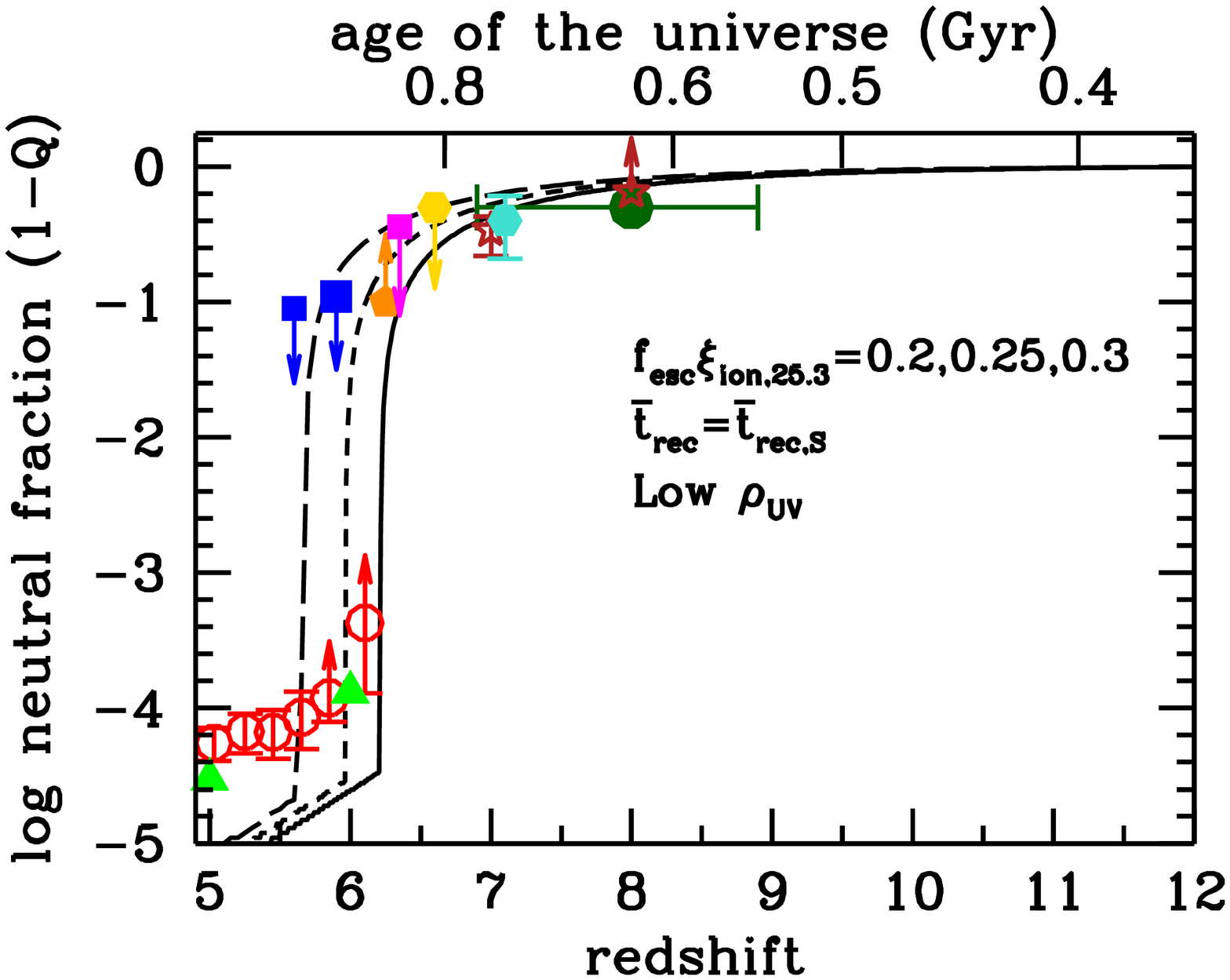}
\includegraphics*[width=0.49\textwidth]{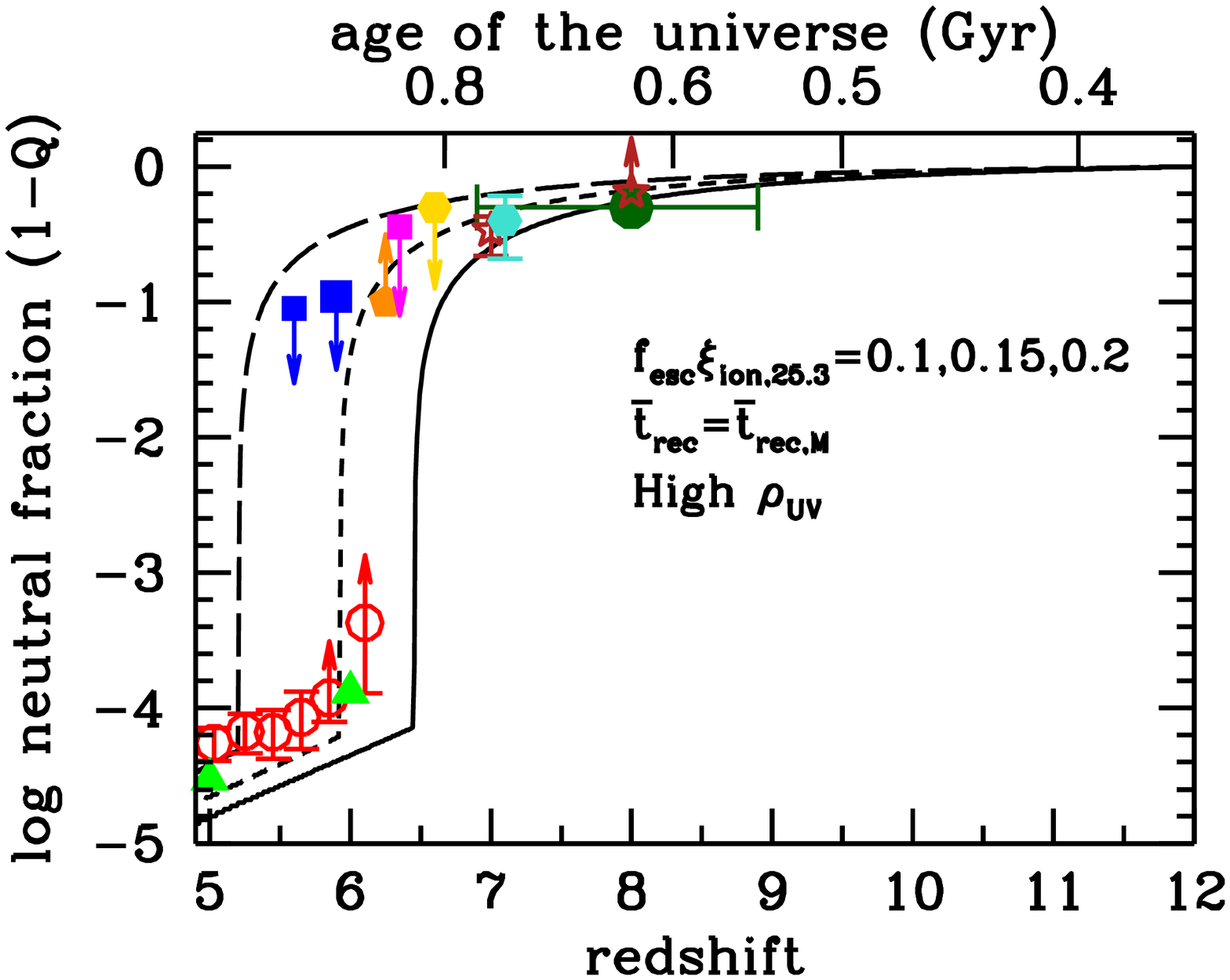}
\includegraphics*[width=0.49\textwidth]{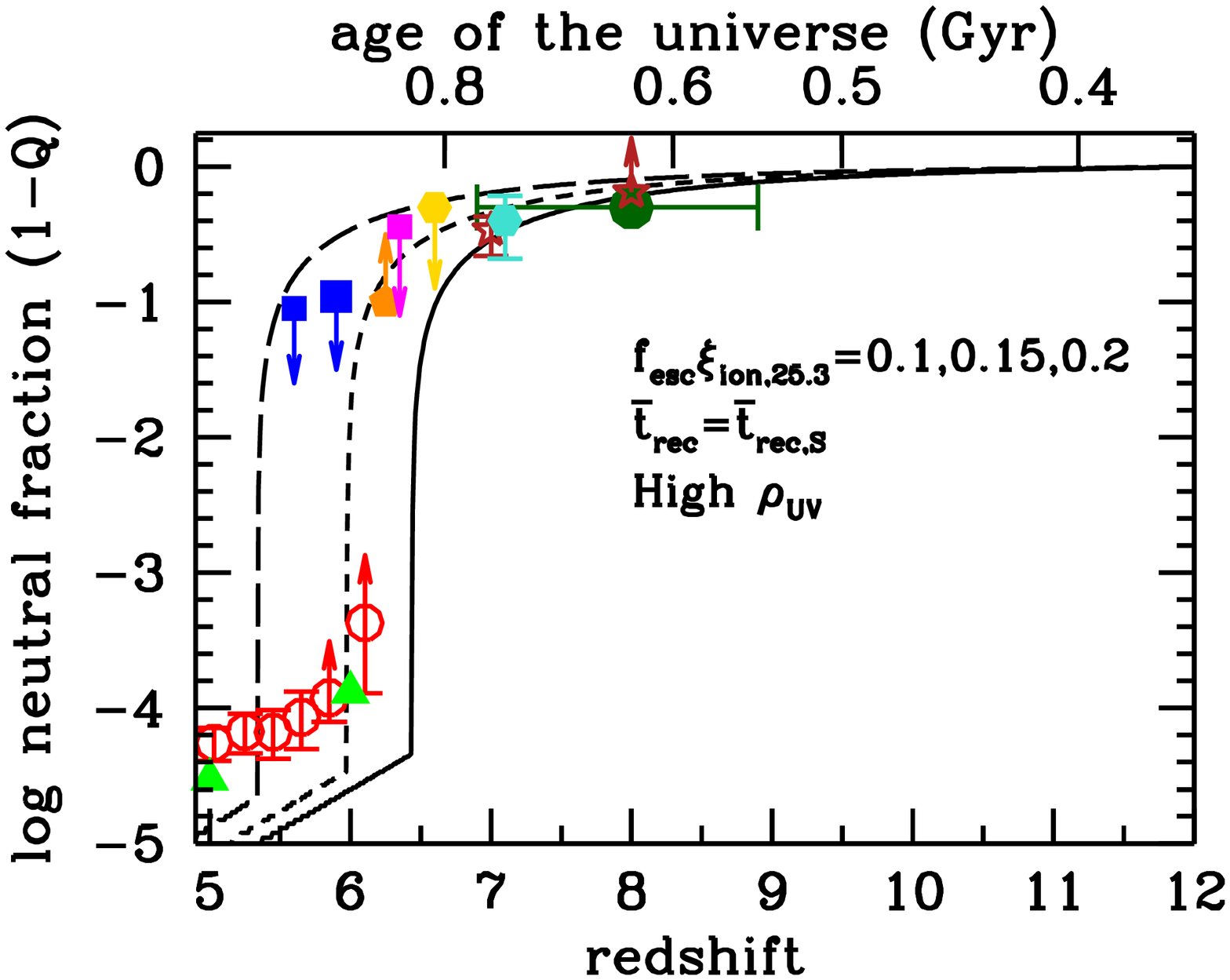}
\caption{Star-forming galaxies as sources of reionization. The reionization histories predicted by numerically integrating Equation
(\ref{eq:dQdt}) from $z=12$ onwards are shown for two models (``high" and ``low") of the  galaxy comoving UV luminosity density and two
different expressions of the IGM recombination timescale. The data points are the same of Figure \ref{fig1}.
Top left panel: ``high $\rho_{\rm UV}$" model with $\bar t_{\rm rec}=\bar t_{\rm rec,M}$ and $f_{\rm esc}\xi_{\rm ion,25.3}=0.2,0.25,0.3$.
Top right panel: same with $\bar t_{\rm rec}=\bar t_{\rm rec,S}$.
Bottom left panel: ``low $\rho_{\rm UV}$" model with $\bar t_{\rm rec}=\bar t_{\rm rec,M}$ and $f_{\rm esc}\xi_{\rm ion,25.3}=0.1,0.15,0.2$.
Bottom right panel: same with $\bar t_{\rm rec}=\bar t_{\rm rec,S}$.
In all panels, the constant LyC photon return into the IGM, $f_{\rm esc}\xi_{\rm ion,25.3}$, is expressed in units of $10^{25.3}\,\xiunits$.
All histories are consistent within the errors with the Thomson scattering optical depth measured by the {\it Planck} satellite \citep{planck16}.
Note how, in the ``high $\rho_{\rm UV}$" solution, values of $f_{\rm esc}\xi_{\rm ion,25.3}\gta 0.15$ are required for
overlap to occur at $z\gta 5.9$. The constraint becomes $f_{\rm esc}\xi_{\rm ion,25.3}\gta 0.25$ in the ``low $\rho_{\rm UV}$" case.
Large LyC photon returns do not reproduce the neutral hydrogen fraction observed in the post-overlap IGM at $5<z<6$.
}
\label{fig3}
\end{figure*}

\subsection{Pre-Overlap}

We can now use our revised reionization equation to bridge the transition from the pre-overlap to the post-overlap era,  
and ask, e.g., what values of the LyC photon return $f_{\rm esc}\xi_{\rm ion}$ may, in combination with a given galaxy UV emissivity, 
reproduce all the observational constraints on the ionization state of the $z>5$ universe. 
We have therefore integrated Equation (\ref{eq:dQdt}) numerically from redshift $z=12$ onwards, using the two expressions for the galaxy 
luminosity density in Equations (\ref{eq:rho16}) and (\ref{eq:rho13}) together with the two different recombination timescales in an attempt to bracket the uncertainties. 
Figure \ref{fig3} shows the resulting reionization histories obtained by assuming a {\it constant} LyC photon return into the IGM, $f_{\rm esc}\xi_{\rm ion,25.3}=0.1,0.15,0.20$
in the ``high $\rho_{\rm UV}$" case, and $f_{\rm esc}\xi_{\rm ion,25.3}=0.2,0.25,0.30$ in the ``low $\rho_{\rm UV}$" case, with higher photon returns naturally leading 
to earlier epochs of overlap. All these histories are consistent within the errors with the Thomson scattering optical depth measured by 
the {\it Planck} satellite. In the ``high $\rho_{\rm UV}$" solution, values of $f_{\rm esc}\xi_{\rm ion,25.3}\gta 0.15$ are required for 
overlap to occur at $z\gta 5.9$, while the constraint becomes $f_{\rm esc}\xi_{\rm ion,25.3}\gta 0.25$ in the ``low $\rho_{\rm UV}$" case. 
Note how, even for photon production efficiencies as high as $\xi_{\rm ion}=10^{25.5}\,\xiunits$ \citep{stanway16,matthee17}, relative photon leakages into the IGM 
of $f_{\rm esc}\gta$10\% appear to be needed. These escape fractions are in qualitative agreement with many recent studies of galaxy-dominated reionization 
\citep[e.g,][]{ishigaki17,gnedin16,khaire16,mitra15,bouwens15planck,robertson15}. 

As mentioned in the previous section, the kSZ effect provides an upper limit to the duration of reionization. All histories in Figure \ref{fig3} with $z_{\rm end}\ge 5.9$   
are characterized by $\Delta z\equiv z_{10}-z_{\rm end}<4$, which is consistent with current constraints from {\it Planck} \citep{planck16}. 
Reionization therefore proceeds rather quickly: the shortest durations ($\Delta z\lta 3.3$) are found in the ``low $\rho_{\rm UV}$" solutions with $\bar t_{\rm rec}=\bar t_{\rm rec,S}$. 
Low luminosity density models typically have lower redshifts of reionization, with $z_{50}=$ 6.9--7.3, compared to ``high $\rho_{\rm UV}$" models, with $z_{50}=$ 7.1--7.8. 

\subsection{Uncertainties and Limitations}

Cosmic reionization involves a complex interplay between the abundance, clustering, spectrum, and leakage of LyC radiation of photoionizing sources,
and the density, clumpiness, temperature, and spatial structure of intergalactic gas, and it is likely that some of the assumptions adopted in our 
numerical integration of the reionization equation may not be fulfilled in practice. In the previous section we postulated, for example, that normal, star-forming galaxies
dominate the production of LyC photons in the pre-overlap, overlap, and immediate post-overlap stages of reionization with a constant ionizing photon return into the
IGM. Figure \ref{fig3} shows, however, that large, redshift-independent LyC photon returns may be difficult to reconcile with the neutral hydrogen fraction observed in the post-overlap IGM. 
In the ``high $\rho_{\rm UV}$" model, in particular, star-forming galaxies can be the source of the cosmic ionizing emissivity at redshift 5 if $f_{\rm esc}\xi_{\rm ion,25.3}\simeq 0.04$ 
(for $\bar t_{\rm rec}=\bar t_{\rm rec,S}$), a value that is too low for overlap to occur at $z\gta 5.9$. 

Nearly all models with constant LyC photon returns into the IGM do not recombine rapidly enough at $5<z<6$ to reproduce   
the large Gunn-Peterson optical depths observed in the post-overlap universe. And while a number of poorly-known input parameters may affect our calculations at these epochs, 
from the opacity of the LLSs through the spectrum of ionizing sources to the temperature of the recombining IGM, it is likely that a time-varying photon return 
may be key for a detailed modeling of the reionization era \citep[see, e.g.,][]{haardt12,kuhlen12,robertson13,khaire16,sharma17}. 
This may be associated with an evolving globally-average escape fraction -- with galaxies at $z\gta 6$ being more porous than their lower redshift counterparts -- and/or 
with an increasing LyC photon production efficiency with redshift. The latter has been shown to be as large as $\xi_{\rm 
ion}=10^{25.5}\,\xiunits$ in stellar population synthesis models at low metallicity that include the impact of binary stars \citep[][]{stanway16} and in observations of luminous 
LAEs at $z\sim$6-7 \citep{stark15,matthee17}. A rise in the effective photon production efficiency may also be caused by additional sources of UV radiation at early epochs other than massive
stars, like, e.g., active galactic nuclei (AGNs) \citep[e.g.,][]{giallongo15,madau15,chardin17,laporte17}.
We also note that, at redshifts $8<z<12$, our extrapolated galaxy UV emissivity is plagued by uncertainties, and does not consider the possibility of a rapidly declining luminosity density 
in the 260 Myr from $z=8$ to $z=12$ \citep[e.g.,][]{oesch15}. The inclusion of such a downturn would only strengthen the case for a late reionization epoch.

At a more basic level, we have integrated the revised reionization equation under the assumption that Equation (\ref{mfpLLS}) correctly represents the volume-averaged opacity of LLSs.
In detail, however, direct measurements of the mean free path at $z\gta 5$ may be biased towards higher values by the quasar proximity effect \citep{daloisio16}.
Optically thin, highly-ionized absorbers with hydrogen columns below $N_\nHI<1.6\times 10^{17}\,\cmm$ (``sub-LLSs") may also play a
non-negligible role in limiting the mean free path \citep[e.g.,][]{haardt12}. Rather than in the source term, we have included the contribution of these sub-LLSs
to photon losses in the radiative recombination term. 

We have extrapolated observations of the mean free path at $z<5.5$ to higher redshifts, which may be dangerous if the 
nature of the LLSs changes during reionization, as the ionizing background becomes weaker \citep[see, e.g.,][and references therein]{sobacchi14,munoz16}.
In our model, LLSs only control the mean free path close to and after overlap, while it is the low-density neutral IGM 
that provides the bulk of the opacity at $z\gta 6$. We have checked, for example, that a more rapid decrease with redshift of the LLS mean free path at $z>5.5$ 
(vs. Eq. \ref{mfpLLS}) would have little effect on the predicted volume-averaged neutral fraction at $z\ge 6$.    
Nevertheless, a firmer understanding of the nature and evolution of the LLSs at these epochs seems essential for an accurate modeling of the late stages of reionization. 

Our semi-empirical approach to clumpy IGM absorption produces a sharp overlap phase, a sudden change in the neutral fraction when reionization is almost completed. This is 
clearly visible in Figures \ref{fig1} and \ref{fig3}, and marks the epoch when the photon mean free path becomes determined by the LLSs rather 
than by the typical size of \HII\ bubbles. According to Equation (\ref{mfpLLS}), the mean separation of the LLSs exceeds 40 comoving Mpc (cMpc) at redshift 6. 
Therefore, if the typical size of \HII\ regions remain smaller than 40 cMpc during reionization, it is the diffuse IGM (with a ``swiss-cheese" ionization topology) 
that will control the photon mean free path, and a well-defined overlap epoch should ensue. A dramatic rise in the mean free path (and in the corresponding UV background intensity) 
was already present in the first numerical hydrodynamics simulations of reionization \citep{gnedin00}, and is also seen in radiative transfer simulations that account for 
LLS absorption \citep{gnedin06,gnedin14}. One should be cautious, however, about numerical artifacts of studying reionization in simulations of finite box sizes that are 
comparable or smaller than the mean free path of LyC radiation at $z\lta 6$. A more gradual transition from bubble-dominated ionization at early times to a 
smoother ``web-dominated" topology characteristic of the post-reionization universe has been argued for by, e.g., \citet{furlanetto09}.
And while a rapid decrease in the volume-weighted neutral hydrogen fraction appear consistent with a number of recent observational constraints on the
ionization state of the $z=$ 5--9 universe, we should recognize that many of them are just educated guesses that rely on constantly changing information.

The resolution of many of the uncertanties discussed above feels overdue and may soon be achieved by the large wavelength coverage, unique sensitivity, and spectroscopic and imaging capabilities 
of the James Webb Space Telescope together with ongoing and future experiments aimed at measuring the redshifted 21-cm signal from neutral hydrogen during the epoch of reionization.

\section*{Acknowledgements}

We thank R. Bouwens, N. Gnedin, and F. Haardt for many useful discussions on the topics presented here, and the anonymous referee for valuable comments and suggestions. 
Support for this work was provided by NASA through grant HST-AR-13904.001-A (P.M.). P.M. also acknowledges  a NASA contract 
supporting the WFIRST-EXPO Science Investigation Team (15-WFIRST15-0004), administered by GSFC, and thanks the Pr\'{e}fecture 
of the Ile-de-France Region for the award of a Blaise Pascal International Research Chair, managed by the Fondation de l'Ecole Normale Sup\'{e}rieure. 

\bibliographystyle{apj}
\bibliography{paper}

\label{lastpage}
\end{document}